\documentclass[vecphys]{svmult}

\usepackage{makeidx}         
\usepackage{graphicx}        

\makeindex             

\def      \comma  {\; , \; \;}

\begin{document}

\title*{Spontaneous Breaking of Space-Time Symmetries}
\author{Eliezer Rabinovici}

\institute{Racah Institute of Physics, The Hebrew University of
Jerusalem, 91904, Israel.  \texttt{eliezer@vms.huji.ac.il}}

%
%
\maketitle

\begin{center}
\textbf{Abstract} \footnote{Based on the  contribution to "String Theory and Fundamental \\ Interactions.
Celebrating Gabriele Veneziano on his 65th Birthday", edited by M. Gasperini and J. Maharana,
to be published by Springer, \\ and Lecture given at   Les Houches summer school:  "String Theory and The Real World From Particle Physics to Astrophysics",     Session LXXXVII, 2007.}
\end{center}
{\small Kinematical and dynamical mechanisms leading to the
spontaneous breaking of space-time symmetries are described. The
symmetries affected are space and time translations, space
rotations , scale and conformal transformations.  Applications are
made  to solidification, string theory compactifications, the
analysis of stable theories with no ground states, supersymmetry
breaking  and the determination of the value of the vacuum
energy.}

\newpage

\section{Introduction}


This being a contribution to honor Gabriele Veneziano, I allow
myself to open with some personal words. I have first heard Gabriele's
name on the radio when the late Yuval Ne'eman described  the great
importance of young Gabriele's work.  That was in the late
sixties, several years later as a student I had the privilege to
learn from a still very young Gabriele  about  the dual model in
full detail. These were outstanding lectures. Over the years I
have learned many more things from Gabriele some of it through
direct collaborations, in parallel we had developed a personal
friendship for which I am grateful.

 It is not uncommon for young scientists  to complain that their teachers didn't
educate them appropriately and did not really pass them/point them
to  the relevant  information. I may have some such complaints of
my own  but not to Gabriele.  I would have liked for example to
know earlier  about the ideas of Kaluza and Klein. So in order to
somewhat reduce  the complaints that will be directed at me, I
would like to use this opportunity to describe something that it
is not taught extensively in particle physics courses,  mechanisms
to spontaneously break space-time symmetries. The world around us
is actually neither  explicitly  invariant under  translations,
nor under rotations. It also is not explicitly invariant under
scale and conformal symmetries. In this work we will review
various mechanisms to break all these space-time symmetries.  I
think they may yet play an important role in particle physics as
well. I will first describe attempts to break translational
invariance kinematically by imposing specific boundary conditions.
Then I will review the Landau theory of solidification and an
attempt to apply it to generate a dynamical mechanism for
compactifications. I will discuss both the success and challenges
of that approach. Next, in the context of breaking time
translational invariance, I will discuss various systems which are
well defined but have no ground state. Following a review of the
breaking of scale invariance and conformal invariance,  I will
also not miss this opportunity to describe in a Katoish manner
that the vacuum energy in conformal/scale invariant theories is
very constrained, and its zero value does not depend on the
presence or absence of any spontaneously generated scales. This
may eventually be recognized as an important ingredient in
understanding and explaining the cosmological constant problem.

\newpage

\section{Spontaneous Breaking of Space Symmetries}

\vspace{.5 cm}

Space symmetries include space translations and space rotations,
we  address the spontaneous breaking of these space symmetries.
This occurs for example when a liquid solidifies and a lattice is
formed.
 The standard manner to identify the ground state of a system is to construct what is called the effective potential.
  The symmetry properties  of the ground state determine whether  a spontaneous breaking
 of  symmetries  which are manifest in the Lagrangian occurs.

Let's review the manner in which the effective potential is
constructed. One first considers all wave functionals which have
the same expectation value of the field operator $\hat{\phi}$

 \begin{equation}
 <\Psi(\phi )|\hat{\phi} |\Psi(\phi )> = \tilde{\phi} \ .
 \end{equation}

Out of these subset of wave functionals, one chooses that
particular wave functional which minimizes the expectation value
of the Hamiltonian. One calls it $V_{eff}(< \phi >)$

\begin{equation}
 V_{eff}(\tilde{\phi}) =
min_{\tilde{\phi}}<\Psi(\phi)|\hat{H}|\Psi(\phi)>.
\end{equation}

  Eventually one draws a picture portraying
 $V_{eff}$ as a function of $\tilde{\phi}$ and one searches for its
minimum. The wave functional for which this  energy minimum was
obtained is the wave functional of the ground state of the system.
However, one usually ignores the possibility that the ground state
wave function would correspond to a non-constant(in $x$)
expectation value $<\phi(x)>$.  Of course it makes much easier the
drawing of pictures in books, here however we will discuss cases
where $<\phi(x)>$ actually does depend on $x$ when evaluated in
the ground state.

Why does one usually only consider wave functionals with constant
values of  $<\phi(x)>$?



 The reason is expediency,  when one wants
 to pick up the ground state of the system among various candidates, one
 is  interested only in the winner, that is the true
 ground state. One does not  care about missing out candidate states whose energies
 are just above that of the ground state of the system. As
 generally   spontaneous breakdown of
 space-time symmetries in the ground state is not expected, one considers it enough
to search for the ground state only among those candidates for
which  $<\phi(x)>$
 is constant. However that need not always be the case.

\subsubsection{ Kinematics:  Attempts to Break  Spatial Translational Invariance Through Boundary Conditions }
\vspace{.5 cm}

 I will first describe an easy way to attempt to break space symmetries.
 That is to break the symmetries not by the dynamics of the  system but kinematically,
  by imposing certain boundary conditions. This easy
 solution, is a mirror to what is done in String Theory in several
 cases,
 including when one is considering  brane sectors. To try and
 break translational invariance by boundary conditions one
 considers for example a system which depends on a scalar field $\phi$.
  Assume the system lives in a box  extending from $-L$ to $L$,
  and impose the condition of anti-periodicity, namely
  \begin{equation}
\phi(L) = -\phi(-L),
 \end{equation}

 where $L$ is the spacial cutoff we put on the system.

\begin{figure}
\begin{center}
\includegraphics [scale=2, height=7cm]{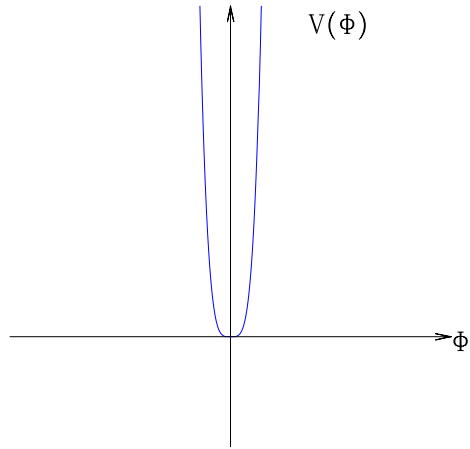}\\
\caption{Unbroken symmetry} \label{unbroken}
\end{center}
\end{figure}

If the system at hand is described by an effective
  potential that has only one minimum (see fig. \ref{unbroken}),

 where there the expectation value $<\phi>$ vanishes, then there is
  no effect resulting from  imposing the boundary conditions. The ground state does fulfill
  the boundary condition, and it
   remains the one which does not break translational
  invariance.  From the point of view of the wave functional, it is
  concentrated around $\phi = 0$.

\begin{figure}
\begin{center}
\includegraphics [scale=2, height=7cm]{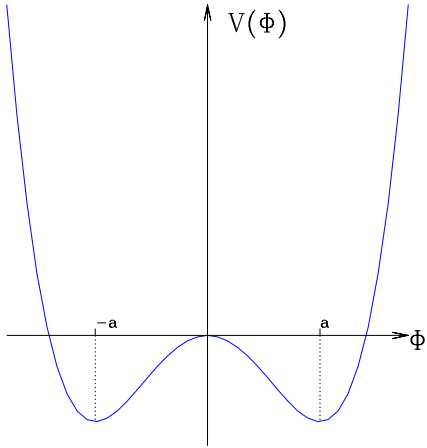}\\
\caption{Broken symmetry} \label{broken}
\end{center}
\end{figure}

 However, consider  the double well potential, (fig. \ref{broken}), (in circumstances where there is
no tunneling).

\begin{figure}
\begin{center}
\includegraphics [scale=2, height=7cm]{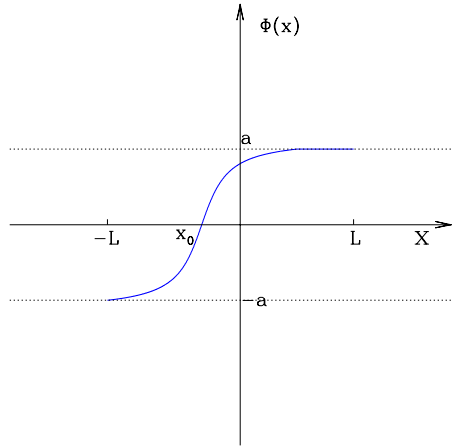}\\
\caption{A soliton attempts to break translational invariance.}
\label{soliton}
\end{center}
\end{figure}

   In this case  the effective potential has two minima, one at $\phi = a$
  and the other at $\phi = -a$. Imposing the boundary condition
  removes  both of the possible true vacua of the system,
  because neither the ground state for which $<\phi> = -a$, nor
  the ground state for which  $<\phi> = a$ obey the boundary
  condition. One is driven to look for another type of
  ground state. We know, for example,  that in a two dimensional system composed only of
  scalar fields there is a finite energy solution, which is a soliton that at $L$ has a
  value $a$ and at $-L$ has a value $-a$, see fig. \ref{soliton}.   An anti-soliton will have
  the opposite values. This is a stable topological configuration,
  and one may imagine that indeed  in such a system there is no
  translational invariance, because the ground state will have to
  be such that its spacial expectation value follows  the values of the
  soliton field, and thus is not translational invariant.

  It is true that by imposing the boundary conditions one has
forced the system  into the
  soliton sector, but one has to remember that this system has a
  zero mode. Technically, if one solves the small fluctuations of
  the scalar field in the presence of a background, which is a
  soliton, one finds that there is a zero mode. This zero mode is
  a reflection of the underlying translational invariance and it actually
  tells  that one is not  able to determine, by energetic considerations,
   where the inflection point $x_{0}$, the point  from which one turns from
   one vacuum to the other, (see fig. \ref{soliton}) occurs. Actually there is a valid soliton solution for each value of $x_{0}$.

 Why is this important? At the case at hand, the zero mode is normalizable.  This
   amounts to saying that the soliton mass is finite. In such a
   case, there is actually no bulk violation of translational
   invariance. What one needs to do is to construct  an eigenstate configuration,
    which is an eigenstate of the linear momentum operation,  a plane wave in terms
    of the center of mass coordinate of the soliton. The lowest energy state which corresponds
   to a momentum state  has $p =0$, one has restored in this way
   translational invariance. The only problem will be to fix the
   system very near the edges, but in the bulk,  the symmetry has been restored and  there is no breaking of bulk  translational invariance.
   Could one still have a case where the boundary conditions do cause a spontaneous breaking of translational invariance?


This may occur when one drives the mass of the soliton to infinity
by an
   appropriate choice of parameters. When the mass of the soliton
   is infinite, physically one cannot form a linear momentum state
   out of it, and technically the zero mode ceases to be
   normalizable. In such a case,  one does indeed break
   translational invariance spontaneously by fixing the point
   where the soliton makes the transition from one vacuum to the
   other. This occurs for example in String
   Theory in a sector containing infinite branes, branes which have finite energy
   do not break translational invariance, and one can build out of
   them linear momentum states. However, branes which extend up to
   infinity  carry infinite energy, and therefore  do lead to the
   breakdown of translational invariance.

   I will mention  at
   this point that once upon the time,  when people were
   considering the breakdown of extended global supersymmetries,
   there was a predominant common wisdom which claimed that one  cannot
   break down extended global supersymmetry to anything but $\mathcal{N} =
   0$. That is either all the supersymmetries are manifest
   together, or they are all broken  together. The argument
   went in the following way:
   one writes the formula for the
   Hamiltonian

 \begin{equation}
  H = \sum_{\ualpha} \bar{Q}^{I}_{\ualpha } Q^{I}_{\ualpha },
 \end{equation}

 where $I = 1,\dots, \mathcal{N}$ is \emph{not} summed, it is a non-trivial
 constraint to get the same Hamiltonian  by summing over different
 supersymmetry generators. When one can do that one has an extended
 supersymmetry. However it is clear from this that if the Hamiltonian does not vanish on the
ground state, then  some of the $Q^{I}$ (for each $I$
independently) do not annihilate the ground state. Therefore the
supersymmetries are either all preserved or all broken.

 This type of argument assumed implicitly  that Poincar\`e
invariance is present in the system. If one now considers other systems, (see for example \cite{Fubini:1984hf}, \cite{Hughes:1986dn}), where part of
the Poincar\`e invariance is preserved and part is broken, this
exposes a loop hole in the former argument.  In the absence of
full translational invariance(due to the presence of infinite mass
branes) one may obtain fractional BPS states, and one may
break $\mathcal{N} = 4$ down to $\mathcal{N} = 2$, $\mathcal{N} =2$
to $\mathcal{N} =1$ and various other combinations.

 This is an example where spontaneous breaking
of translation invariance occurs, it has an impact also on the
partial breaking of global supersymmetry and, if one wishes, this
is a way to break translation invariance by forcing the system,
using boundary conditions, to a certain super-selection sector.

This is not what I mainly want to discuss here. I would like to
discuss a situation where the dynamics of the system drives the
spontaneous breaking of translational and rotational invariance.

\vspace{1 cm}

\subsubsection{Dynamics: The  Landau  Theory of Liquid-Solid  Phase Transitions}

\vspace{.5 cm}

\begin{figure}
\begin{center}
\includegraphics [scale=2, height=5cm]{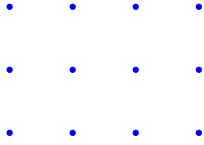}\\
\caption{The solid lattice breaks most of the translational and rotational invariances} \label{lattice}
\end{center}
\end{figure}

Let us now turn to discuss the transition between a liquid and a
solid. This follows the seminal work of Landau \cite{Landau}. In a
monumental paper, he described  spontaneous symmetry breaking of
both, internal and  space-time symmetries. Consider  a liquid,
 a system whose lagrangian is either relativistic or
non-relativistic, and it possesses full rotational and
translational invariance.  On the other hand, a solid is a system
which maintains only a very small part of the translational
invariance and rotational invariance, (fig. \ref{lattice}).

Let us simplify the study by ignoring the point structure  at each
lattice point which a solid may have. That is let's not consider
the atomic structure at each point.  One focuses first  on  the
question of how does the simplest lattice forms.

 I will describe this following Landau and then, following
 \cite{EFR}, I am going to describe applications to
String Theory. Landau starts by defining the Landau order
parameter to monitor  the transition between a solid and a liquid.
It is a scalar order parameter  $\varrho(\vec{x})$

 \begin{equation}
  \varrho(\vec{x}) = \varrho_{s}(\vec{x}) -\varrho_{0},
 \end{equation}

 the difference between the non-translational non-rotational
invariant density of the solid $\varrho_{s}(\vec{x})$,  and the
constant density $\varrho_{0}$ of the liquid. Next, consider the
Fourier decomposition of $\varrho(\vec{x})$

\begin{equation}
\varrho(\vec{x}) = \sum \varrho(\vec{q}) e^{\I \vec{q} \cdot
\vec{x}} + h.c. \ .
\end{equation}

It is useful to use as order parameters the Fourier components
$\varrho(\vec{q})$ .

The question is thus: Does the wave functional of the ground state
have support on $\vec{q} \ne \vec{0}$?  If the answer is positive,
then at the very least continuous translational  space symmetry
would be spontaneously broken. This will be determined by studying
the Landau-Ginsburg effective action as expressed in terms of the
order parameter $\varrho(\vec{q})$. The first relevant term of the
Landau-Ginburg action is quadratic in the order parameter and is
given by:

\begin{equation} \mathcal{L}_{0} = \int \D \vec{q}_{1} d \vec{q}_{2}
 \varrho (\vec{q}_{1}) \varrho ( \vec{q}_{2}) A(|\vec{q}_{1}|^{2})\udelta ( \vec{q}_{1} +\vec{q}_{2}).
\end{equation}

The delta function $ \udelta(\vec{q}_{1}+\vec{q}_{2})$ enforces
translational invariance, while rotational invariance is preserved
by the dependence on $|\vec{q}|^{2}$ of the function
$A(|\vec{q}|^{2})$. The function $A(|\vec{q}|^{2})$, like in any
Landau-Ginsburg potential, is determined by the microscopic
theory. In the particular case at hand, it will depend on the
hardcore potential component in the atoms involved  and on other
possible potentials, as well as on the temperature of the system.
In the case of neutron stars, studied in \cite{Baym}, the Pauli
exclusion principle plays a role  in determining the function
$A(|\vec{q}|^{2})$.

 Let us treat first an example that we are familiar with,  that of a
free massive spin-zero particle in a relativistic field theory.
        In that case the function $A(|\vec{q}|^{2})$ is

\begin{equation} A(|\vec{q}|^{2}) = |\vec{q}|^{2} + m^{2}. \end{equation}

\begin{figure}
\begin{center}
\includegraphics [scale=2, height=7cm]{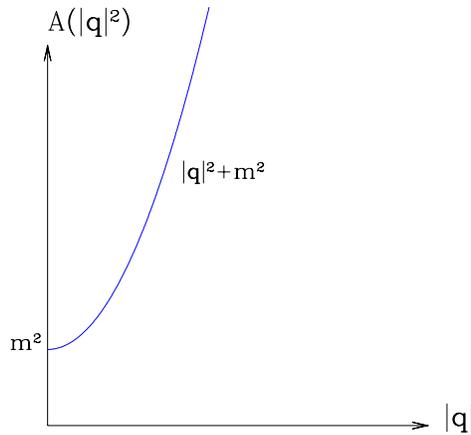}\\
\caption{The form of $A(|\vec{q}|^2 )$ in a free massive
relativistic field theory does not lead to spontaneous breaking of
translational invariance.} \label{free}
\end{center}
\end{figure}

 This  has a minimum at $|\vec{q}|^{2} = 0$, as shown in fig. \ref{free},  and thus the function
$\varrho(\vec{q})$ should get the support only at
$\vec{q}=\vec{0}$,
 there is no spontaneous breakdown of translational invariance
in this case.

 \vspace{.7 cm}

\begin{figure}
\begin{center}
\includegraphics [scale=2, height=7cm]{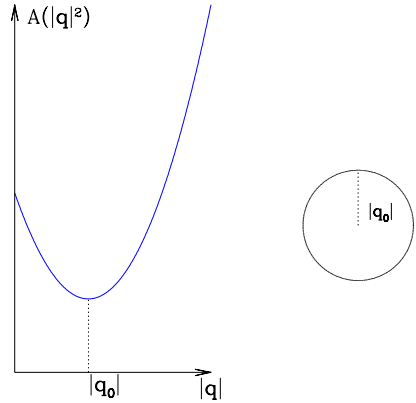}\\
\caption{Example of a function $A(|\vec{q}^{2}|)$ which leads to
the breaking of translational invariance. An explicit
microscopical realization of a such a form appears in neutron
stars \cite{Baym}. The wave functional is concentrated at most on
the shell of a sphere of radius $|\vec{q}_{0}|$. }
\label{interacting}
\end{center}
\end{figure}

In the presence of interactions things may become more complicated,
for example I am not familiar even with a proof that the Standard
Model ground state does not violate space-time symmetry, (though
most likely it does not). In any case the microscopic theory may
allow a different function for $A(|\vec{q}|^{2})$.
 In particular, assume that the form of $A(|\vec{q}|^{2})$ is as
 given in fig. \ref{interacting} .  In this case, the function $A(|\vec{q}|^{2})$
 has a minimum at a value $|\vec{q}_{0}|^{2} \ne 0$. In such a
 system  the ground state wave
 functional gives rise to a density concentrated around $|\vec{q}_{0}|^{2} \ne 0$.
In particular, one would expect the support to be concentrated
around a sphere in $\vec{q}$-space, whose radius is
$|\vec{q}_{0}|$. So one is in a situation, given
$A(|\vec{q}|^{2})$ of that form, where one has  a spontaneous
breaking of translational invariance, but not yet also a  breaking
of rotational invariance, which is what is needed to form a solid.
It is good enough to break just translational invariance.

The  ground state density does  depend on $\vec{x}$
\begin{equation} \varrho(\vec{x}) =
\int_{\mathcal{S}_{|\vec{q}_{0}|}} \D \Omega  \ \varrho(\vec{q})
e^{\I \vec{q} \cdot \vec{x}} + h.c. \ .
\end{equation}

In  this approximation the wave functional of the ground state is
supported on a sphere $\mathcal{S}_{|\vec{q}_{0}|}$ whose radius
is $\vec{q}_{0}$. In particle physics we have become rather
sophisticated, and when one writes down the Landau-Ginsburg
action, one usually requires that the expansion  in the order
parameter to be under control. For example, this means that there
is a limit in which this expansion becomes exact. In the case at
hand, this is not the situation, and it is actually very
complicated, nevertheless one follows the usual Landau-Ginsburg
expansion.

The term which follows the quadratic interaction is a cubic term

\begin{equation} \mathcal{L} = \mathcal{L}_{2} + \mathcal{L}_{3}, \end{equation}

\begin{eqnarray} \mathcal{L}_{3} = \int \D^{3}\vec{q}_{1} \D^{3}\vec{q}_{2}
  \D^{3}\vec{q}_{3} \ \varrho(\vec{q}_{1}) \varrho(\vec{q}_{2})
\varrho(\vec{q}_{3}) \udelta(\vec{q}_{1} + \vec{q}_{2} +
\vec{q}_{3}) \times  \nonumber \\
B(|\vec{q}_{1}|^{2},|\vec{q}_{2}|^{2},|\vec{q}_{3}|^{2},\vec{q}_{1}\cdot
\vec{q}_{2},  \vec{q}_{1}\cdot \vec{q}_{3}, \vec{q}_{2}\cdot
\vec{q}_{3}    ) \ \label{cubic}
\end{eqnarray}

  I am going to assume for the purpose of illustration, as
Landau did, that this is a good perturbation, namely that when one
considers $\mathcal{L}_{3}$ one is going already to assume that
the support of $\varrho$ comes from only those values of $\vec{q}$
such that $ |\vec{q}_{1}|^{2} \cong
|\vec{q}_{2}|^{2}\cong|\vec{q}_{3}|^{2} \cong
 |\vec{q}_{0}|^{2}$. This was determined by $\mathcal{L}_{2}$.

In (\ref{cubic}), once again, the delta function
$\udelta(\vec{q}_{1} + \vec{q}_{2} + \vec{q}_{3})$
 enforces the explicit translational invariance, and the dependence of $B$
on the momentum respects both translational and rotational
invariance.

The integral in the $\vec{q}$`s  is not over all possible values,
but only over those whose lengths is determined by
$|\vec{q}_{0}|^{2}$, which in turn was fixed by $\mathcal{L}_{2}$.

\begin{figure}
\begin{center}
\includegraphics [scale=2, height=7cm]{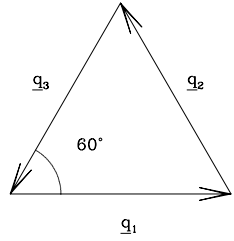}\\
\caption{The sphere $\mathcal{S}_{|\vec{q}_{0}|}$ is triangulated
due to the presence of a cubic term in the Lagrangian. Since in
this approximation all the sides of the triangles have the same
length, their angles are determined to be $60^{o}$. Rotational
invariance is thus spontaneously broken.} \label{triangles}
\end{center}
\end{figure}

An additional structure emerges due to the effect of  the delta
function $\udelta(\vec{q}_{1} + \vec{q}_{2} + \vec{q}_{3})$. It
restricts the candidates for the ground state, to have support on
at least three different values for the $\vec{q}_{i}$. The three
vectors appearing need to sum up to give a triangle, see fig
\ref{triangles}.

Actually they are six if the field is real since one needs

\begin{equation} \varrho(\vec{q}) = \varrho(- \vec{q}). \end{equation}

 Thus one has at least six components of $\varrho(\vec{q})$ which do not vanish.  In
general, instead of  $\varrho(\vec{q})$ having support on
all values of a sphere, they are now broken into triplets where
the $\vec{q}_{i}$ have to sum together to form  triangles, (fig.
\ref{triangles}). In this manner  also rotational invariance is spontaneously broken.

Let's be even more explicit, because we have used the
approximation that all the $\vec{q}_{i}$ have the same length, the
$\vec{q}_{i}$ that tessellate the sphere, have to form equilateral
triangles, as in fig. \ref{triangles}. Equilateral triangles
single out a specific angle $60^{o}$, that is a spontaneous
breaking of rotational invariance. One has obtained a non zero value for
$\vec{q}$ , and one has derived that the ground state is
built out of objects which have to sum up to form triangles
which are equilateral and thus have a $60^{o}$ angle.

 From energetic and combinatorial
considerations, one finds that to be on an extremum, one needs all
the values of $\varrho(\vec{q}_{i})$ to be equal

\begin{equation} |\varrho(\vec{q}_{i})|^{2} = |\varrho(\vec{q}_{0})|^{2}, \end{equation}

which leads to

 \begin{equation} |\varrho(\vec{x})|^{2} = n|\varrho(\vec{q}_{0})|^{2}. \end{equation}

There are a couple of general way to  distribute the triplets, one
in which each $\vec{q}_{i}$ appears in only one of the triplets,
and another in which each value of $\vec{q}_i$ does participate in
two triplets. The number of elements is proportional to $n$ in
both cases, being either $2n/3$ or $4n/3$.  When one does the
analysis, and one estimates the value of $\mathcal{L}_{3}$, one
finds that it decreases as  the  inverse of $\sqrt{n}$

\begin{equation}
 \mathcal{L}_{3} \sim \frac{|\varrho(\vec{q}_{0})|^{2}}{\sqrt{n}}.
\end{equation}

Thus the ground state will be obtained for some finite value of
$n$.  One  needs to consider only a finite number of  triplet
configurations when one searches for the  extrema of the free
energy.  Just three, i.e  six participants   lead to the
following density distribution

\begin{equation} \varrho(x,y) =  \pm \left(\frac{2}{3}\right)^{1/2}  \varrho_{q_{0}}
\left[ \cos(q_{0}x) + 2 \cos \left(\frac{1}{2}q_{0}x \right) \cos
\left(\frac{\sqrt{3}}{2}q_{0}y \right)\right] \ . \end{equation}

 The corresponding free energy is

\begin{equation} \mathcal{L}^{n = 3}_{3} = \frac{2B
\varrho^{3}_{q_{0}}}{3\sqrt{3}}. \end{equation}

For the case of two spatial dimensions  it turns out that if
$\varrho(q_{0}) > 0$ it is
  advantageous to form a triangular lattice, while if   $\varrho(q_{0}) < 0$,
   the dual lattice, which is a honeycomb lattice, is formed.

This required only studying  the minimal possible configuration.
In three spatial dimensions, this would be a candidate for  a two
dimensional lattice in three dimensions, if
 one wishes some type of compactification.

\vspace{.5 cm}

 In three dimensions one needs to consider also larger configurations to obtain the extrema.
 The next candidate  configuration  has six  ($n = 6$), i.e
  twelve values of $\vec{q}$. This is  a more complicated
  configuration, whose density distribution is

  \begin{eqnarray}
\varrho(x,y,z) &=&
\frac{2}{\sqrt{3}}\varrho_{q_0}\bigg[\cos\left(\frac{\sqrt{2}}{2}q_0
x\right) \cos\left(\frac{\sqrt{2}}{2}q_0 y \right) +
 \cos\left(\frac{\sqrt{2}}{2}q_0 x \right)\cos\left(\frac{\sqrt{2}}{2}q_0 z\right)
+\ \nonumber \\ && \cos\left(\frac{\sqrt{2}}{2}q_0 y
\right)\cos\left(\frac{\sqrt{2}}{2}q_0 z\right) \bigg] \ .
 \end{eqnarray}

   This actually describes  a BCC lattice  (in real space).
The value of $\mathcal{L}_{3}$ is larger than that one of the
former configuration
\begin{equation}
\mathcal{L}^{n=6}_3 = \frac{4B\varrho^{3}_{q_{0}}}{3\sqrt{6}} >
\mathcal{L}^{n=3}_3 \ ,
\end{equation}
and leads to the extrema of  the free energy, being the most
stable configuration.

 From  amazingly simple considerations, one has a prediction  that
solids in three dimensions are all BCC lattices, a very universal
description of the system.
 Before confronting this claim with the data one needs to recall
that the transitions between solids and liquids are not second
order transitions, they are actually first order transitions. So
one  may question  the validity of universality claims in this
context. However, it turns out that in many cases one can arrange
that the solidifications occur as a weak first-order transitions,
in which case approximate  universality properties can be present.

\begin{figure}
\begin{center}
\includegraphics [scale=2, height=7cm]{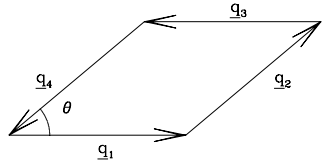}\\
\caption{In the absence of a cubic term, a quartic term would not
suffice classically to induce a spontaneous breaking of rotational
invariance. A rhombus does not single out a preferred angle
$\theta$} \label{rhombus}
\end{center}
\end{figure}

Returning to the data  and following \cite{Alexander},  one
discovers that   about 40 metals, which are on the left of the
periodic table, ( excluding Magnesium (Mg)),  form near the
solidification point  a BCC configuration.

 I will repeat  the
difficulties of the analysis and the argumentations to proceed
with it. The transition is first order - the fact that in many
cases it is a weak first order transition softens this problem.
There is no true expansion parameter in the problem. The
microscopic theory constructing $A$ and $B$ is very
phenomenological , and therefore the real relative stability of
the metal is a very delicate matter.  Even taking all this into a
account  the result and its agreement with a large body of the
experimental data is striking.

Consider what would have happened without a cubic term. In that
case, the term following the quadratic term would be
$\mathcal{L}_{4}$, which schematically would assume the form

\begin{eqnarray} \mathcal{L}_{4} = \int \D \vec{q}_{1}   \D \vec{q}_{2}
\D  \vec{q}_{3}   \D \vec{q}_{4} \ \udelta(\vec{q}_{1}
+\vec{q}_{2} + \vec{q}_{3}  + \vec{q}_{4}) \times \nonumber \\
C(|\vec{q}_{1}|^{2}, |\vec{q}_{2}|^{2} |\vec{q}_{3}|^{2},
|\vec{q}_{4}|^{2}, \vec{q}_{1}\cdot\vec{q}_{2},
\vec{q}_{1}\cdot\vec{q}_{3}, \dots ) \ ,
 \end{eqnarray}

where the delta function enforces translational invariance, and
 the function $C$ should be build by such invariants that maintain both
rotational and translational invariance.

This  does not break rotational invariance, because unlike the
case of  triangles,  the configurations which are enforced now,
assuming perturbation theory, are those of quadrilaterals with
equal sides. But for a  rhombus (fig. \ref{rhombus})  no preferred
angle is singled out.  The rotational invariance is not broken.
Fortunately there is no microscopic symmetry consideration that
rules out the cubic term.

Another interesting type of lattices are the Abrikosov lattices
formed of vortexes, which we do not discuss here.

\subsection{String Theory Compactifications}


 What has been described above has a very solid basis in
nature. What we will describe next is of  a much more speculative
nature, and it is based  on  work by Elitzur, Forge and myself
 \cite{EFR} , in which we try to address the issues of
compactification in String Theory. There are several attitudes one
might adopt  regarding compactification. One, which makes a lot of
sense, is to say that the Universe starts up very small, and the
issue of compactification is an issue of explaining why four
dimensions became very large, while the rest of the dimensions
remain small. This is not what I am going to discuss here.

 Here, I  discuss possible dynamical aspects of compactification taking in account some
 of the hints learnt from the case of solid state physics.
 I don't have  much confidence in human imagination when it is totaly detached from reality,
 and  I would hope that many of the hints available in nature will be  useful to understand other phenomena.
   In particle physics one has learned quite a lot
from the dynamics of solid state physics, and statistical
mechanics systems.

Returning to the case at hand, we have just reviewed a system
which has lost most of its rotational and translational
invariance, and we want to see how a similar  thing might happen
in String Theory. One of the key ingredients driving this behavior
is the presence of a bulk tachyon.

There are actually at least three types of tachyons/instabilities
with whom one is familiar right now in String theory. One is the
Bosonic  closed String Theory tachyon. This instability could well be an
incurable one, nevertheless let's try and follow it.

The other types of instabilities, which we will discuss later, are
 an instability in Open String Theory, an open string tachyon,
and also  localized bulk tachyons.

For the moment we  focus on bulk tachyons, which will be one key
ingredient. Due to them it is preferable for a system in String
Theory containing a tachyon to have a support on a non-zero value
of $q^{2}$.  One can see this from the form of the tachyon, whose
vertex operator is the following

\begin{equation}
 T(x) = e^{\I q_{0}x} + h.c.   \ .
\end{equation}

\begin{figure}
\begin{center}
\includegraphics [scale=2, height=5cm]{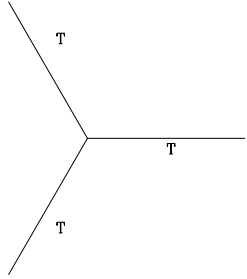}\\
\caption{Tachyonic cubic vertex} \label{vertex}
\end{center}
\end{figure}

To obtain a dimension $(1,1)$ operator, one needs $q_{0}
\ne 0$. Tachyons do give us the starting point that appears in
Landau theory of solidification   (note that here it is not a
minimum consideration). The second key ingredient that we need for
Landau's theory of solidification, in order to obtain not only the
breakdown of translational invariance, but also of rotational
invariance, is the presence of a cubic term. We know from the OPE
(operator-product-expansion) that three tachyons do couple
together, (see fig.\ref{vertex}). In particular,  the OPE between
two tachyons does contain a third tachyon. So we have in a such a
theory a $T^{3}$ term. One indeed has  the necessary ingredients
to try to follow if tachyons could lead to the spontaneous
breaking of rotational and translational invariance in String
Theory, and maybe also to compactification.

\vspace{.5 cm}

In order to be more concrete, we followed the ideas of
\cite{Zamolodchikov}, \cite{Ludwig}   and  tried to handle in a
reliable fashion almost marginal operators.  Consider  a tachyon
which is
 not an exact $(1,1)$ operator, but one which has $q^{2}_{0} = 2 -
\varepsilon $. We will also look at the subset of the full string
background,  a subset which contains  a $c = 2$ sector . We will
not deal here with the question of how the total central charge
remains at the appropriate value, which is zero and how to dress
operators.

As an illustration, consider  the subset of the backgrounds which
are  string moving in flat space, where the piece of the
lagrangian on which we focus is

\begin{equation} \mathcal{L} = \partial X^{1} \bar{\partial} X^{1} +\partial
 X^{2} \bar{\partial} X^{2} + T(X^{1}, X^{2}).
\end{equation}

 From  Landau's theory of solidification, we know that because the
system has support on a $\vec{q}_{0} \ne 0$, and because the free
energy of the system contains  a cubic coupling, we can try to
build the triplet, which again are actually six vectors, so they
get a support in an appropriate way, i.e such that they break
translational and rotational invariance.

 The  $60^{o}$ angle, discussed in the solidification case,
manifests itself in a suggested tachyon configuration:

\begin{equation} T(X^{1}, X^{2}) = \sum_{a=1}^{3} T_{a}
\cos(\sum_{i}^{2}k_{i}^{a} X^{i} ), \label{tachyo} \end{equation}

where the three momenta $\vec{k}^{1}$, $\vec{k}^{2}$, and
$\vec{k}^{3}$ are the following

 \begin{equation} \vec{k}^{1} = k(1,0)
\quad \vec{k}^{2} = k(-1/2, \sqrt{3}/2) \quad \vec{k}^{3} =
k(-1/2, -\sqrt{3}/2).
\end{equation}

\begin{figure}
\begin{center}
\includegraphics [scale=2, height=7cm]{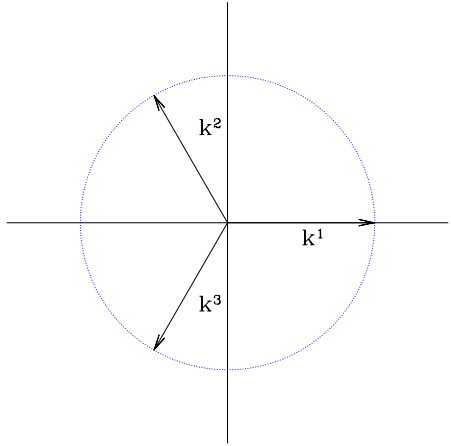}\\
\caption{SU(2) roots} \label{roots}
\end{center}
\end{figure}

All of them  have  $\vec{k}^{2} = 2 - \varepsilon$, and the
structure is very similar to that of the  $SU(3)$ root lattice
(see fig. \ref{roots}), as before for a every $k_{i}$, there is
also the corresponding $-k_{i}$ contribution.

One can simplify  the tachyon potential by taking the ansatz  for
the amplitudes $T_{a} = T$.

 The Lagrangian one needs to solve is the following

\begin{equation} \mathcal{L} = \partial X^{1} \bar{\partial} X^{1} +\partial
 X^{2} \bar{\partial} X^{2} + T(X^{1}, X^{2}),
\end{equation}

and actually one can show that the beta function of the tachyon
alone vanishes to order $\varepsilon$.  So (\ref{tachyo}) is a
solution of the approximate tachyon equations of motion. This
means that had it been up to the tachyon alone one would have
obtained the lattice, perhaps some honeycomb or triangular
lattice, which would break both translational and rotational
invariance. However, this system contains also gravity so one
needs to see what is the influence of the formation of such a
lattice on gravity. As shown in \cite{EFR}, the beta function for
the graviton $\ubeta_{G_{\umu \unu}}$ vanishes (at leading order
in $\ualpha^{'}$) if

\begin{equation} \ubeta_{G_{\umu \unu}} = -R_{\umu \unu} + \nabla_{\umu}
T\nabla_{\unu} T = -R_{\umu \unu} +
\frac{3}{2}\varepsilon^{2}\udelta_{\umu \unu } = 0. \end{equation}

For $D=2$, due to the Liouville theorem  $R_{\umu \unu}$ can be
written as
 $R_{\umu \unu} = a \udelta_{\umu \unu}$, so  one can solve the
 equation by forming a two-dimensional sphere.

 \vspace{.5 cm}

 This is actually a
 highlight of  a model for compactification.  We started by having
 just a tachyon. The tachyon would have produced the lattice
 on its own, but because of the presence of
 the gravity the lattice of tachyons actually causes the
 compactification  of space to a sphere.

However, it turns out, and details are presented in \cite{EFR},
that unfortunately this result is not obtained in a desired
reliable approximation.  The main problem is that in order to do
reliable perturbation theory, we need to do a plane waves
expansion, with the wave lengths representing  a nearly marginal
operator. However, the moment the sphere is formed the topology
changes, and the change of topology means that one should now
expand the fields in terms of spherical harmonics $Y_{l,m}$. This
topological obstruction takes away the reliability of our
calculation.  Some defects may  form in order to resolve this
topological problem, and one conjecture we had at that time was
that actually parafermions, which are defects, form to resolve the
tension. A more complex form of compactification emerges.

 Once again, recall that actually the system, when  fully considered,  has to be coupled to
 the dilaton, in order to maintain  the total central charge.
 According to the Zamolodchikov theorem \cite{Zamolodchikov}, once the system starts to
 flow, the central charge decreases from $2$ and this on its own
 breaks the balance. In a sense, in the case of bulk tachyons
 we were tantalizingly close to obtain an explicit dynamical mechanism for compactification.
However, due to topological obstructions, what was a solution
  for the beta function locally in space,  it cannot be a global
 solution without taking into account other effects.
We will return to the breaking of translational invariance in the
different context of the open string tachyon.

\subsection{Liquid Crystals}

\vspace{.7 cm}

\begin{figure}
\begin{center}
\includegraphics [scale=2, height=3cm]{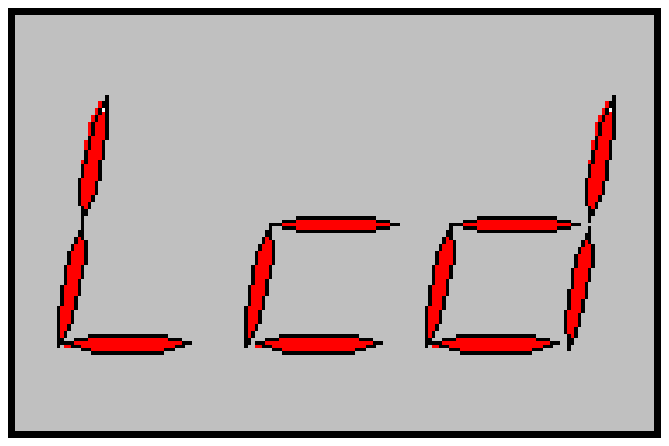}\\
\includegraphics [scale=2, height=4cm]{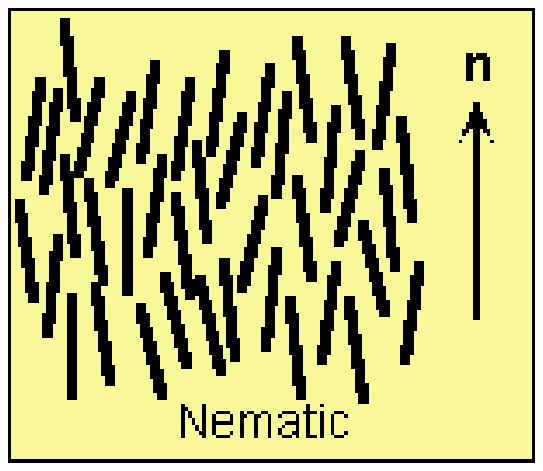}\\
\includegraphics [scale=2, height=4cm]{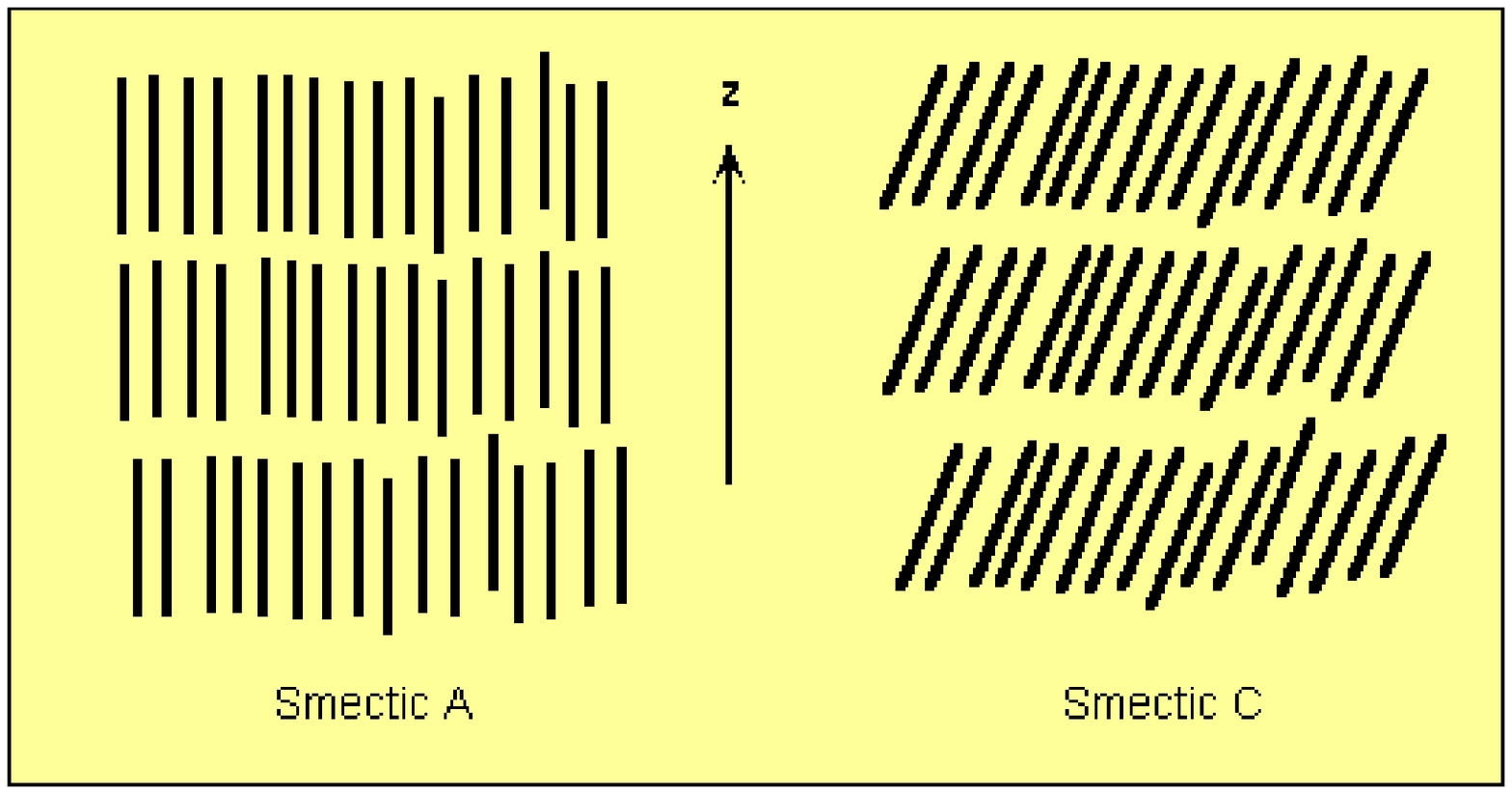}\\
\caption{Various phases of liquid crystals breaking. These systems
exhibit asymmetrical breaking of translational and rotational
invariance.} \label{lcd}
\end{center}
\end{figure}

The tachyon is a scalar order parameter,  String Theory has
additional fields  which carry indexes. In particular, one might
think that  if one looks for a similarity to our universe, maybe
one should would be consider the phase of  liquid crystals. Such
systems are translational invariant in some directions but not in
other (see fig. \ref{lcd}). We will give now examples of that.


There are various types of liquid crystals and one can ask what is
the Landau-Ginsburg theory of them. Actually, one can also ask
about vector potential systems which are described, as gauge
fields are, by vector order parameters. Such systems include
detergents  which posses a hydrophobic and a hydrophilic pole, and
play a crucial role in cleaning our garments.
 One can try   to extract from $\vec{p}(\vec{r})$  the various invariants one wants to use in order to describe
this system, such as $div\vec{p}$,
$curl\vec{p} $, $s_{\ualpha \ubeta} = \partial_{\ualpha}
p_{\ubeta} +
\partial_{\ubeta} p_{\ualpha}$.

 It turns out that one can write
down a Landau-Ginsburg theory for detergents, which explains many
of their very fascinating properties.

 Consider the case of liquid
 crystals, these can also be written
 by choosing  as an order parameter  particular
 spherical harmonic functions.

For illustrative  purposes, we give the dependence of the density
$\phi$ on the angles and on the coordinates\footnote{The index
structure of $\phi$ has been omitted.}

\begin{equation} \phi = \sum_{i} \umu_{i} Y^{2}_{2}(\theta_{i}, \phi_{i})
e^{\I \vec{k}_{i} \cdot \vec{r}_{i} } + h.c. \ . \end{equation}

By assuming the ansatz $\umu_{i} = \umu $, the effective
Landau-Ginsburg free energy is given below

\begin{equation} F \sim (\ualpha_{0} + dk + c k^{2})\umu^{2} - \ubeta \umu^{3} + r
\umu^{4}, \end{equation}

 from which one can extract the properties of nematic, smectic A and
smectic C properties, and many other exciting things for which we
refer to the literature \cite{Alexander}.

\subsection{Boundary Perturbations}

Next I discuss an example where a breakdown of spatial
translational symmetry actually clearly occurs.
 As mentioned above one can formulate an intuitive theorem in the bulk,
 which states that under the renormalization group flow,
  the value of the Virasoro central charge $c$ decreases from its UV  value to a smaller IR value.
This is due to the integrating out of the degrees of freedom and
applies to the unitary sectors of String Theory. In String Theory
with its ghosts  the total central charge vanishes. One can
imagine mechanisms by which the central charge of the ghosts
increases \cite{Elitzur:1998va}, but basically  one needs  to
couple the two systems maintaining a total vanishing central
charge.  This can be done for example with the help of a linear
dilaton and leads to very interesting questions and results.
Generically, the matter central charge will decrease to zero
leaving one with just a $c=0$ topological theory, but there are
also other possibilities. The central charge is related to the
anomaly which exists in the bulk. On the other hand, when one
considers the boundary theory, there are no gravitational
anomalies in it. Thus in that case one can consider tachyonic open
string theory perturbations.
 In the example given in the action below

\begin{equation}
 S = \int_{\Sigma} \mathcal{L}_{CFT} + \int_{\partial \Sigma} g
 \mathcal{O}_{Rel.},
 \end{equation}

 the bulk theory is defined on the surface $\Sigma$ and  on
its boundary $\partial \Sigma$   one adds a relevant operator
$\mathcal{O}_{Rel.}$. There is a boundary  renormalization group
flow, which does not change the bulk central charge and therefore
does not lead to all the problems associated with tachyons in the
bulk.

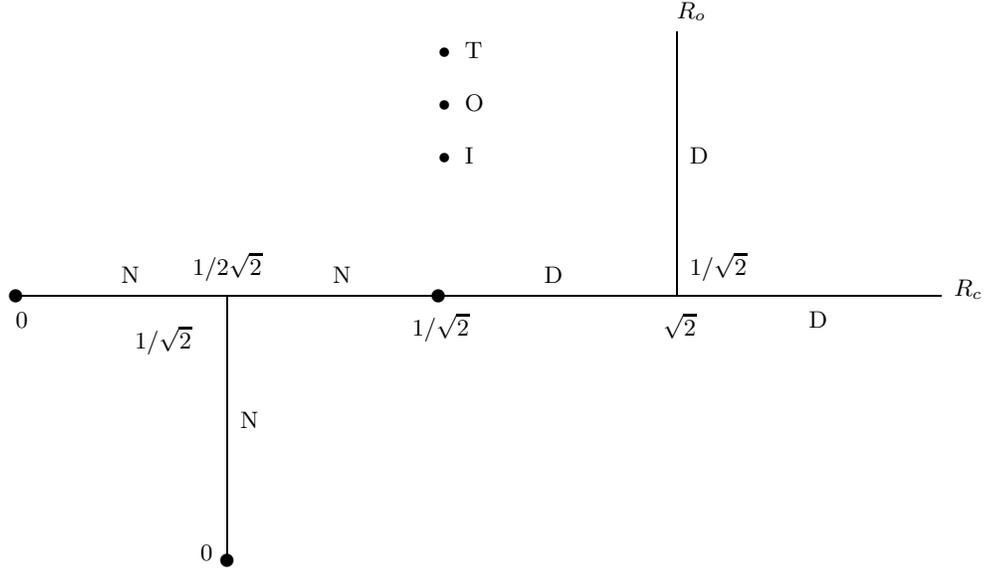
\begin{figure}[h]
\begin{picture}(500,200)(0,-100)
\put(0,0){\circle*{5}} \put(0,0){\line(80,0){350}}
\put(80,0){\line(0,-100){100}} \put(160,0){\circle*{5}}
\put(250,0){\line(0,100){100}} \put(80,-100){\circle*{5}}
\put(40,5){N} \put(85,-50){N} \put(120,5){N} \put(200,5){D}
\put(255,50){D} \put(300,-12){D} \put(355,0){$R_{c}$}
\put(250,105){$R_{o}$} \put(150,-15){$1/\sqrt{2}$}
\put(245,-15){$\sqrt{2}$} \put(67,8){$1/2\sqrt{2}$}
\put(45,-20){$1/\sqrt{2}$} \put(255,8){$1/\sqrt{2}$}
\put(160,50){$\bullet$} \put(170,50){I} \put(160,70){$\bullet$}
\put(170,70){O} \put(160,90){$\bullet$} \put(170,90){T}
\put(0,-12){0} \put(70,-100){0}
\end{picture}
\vspace{2cm} \caption{Map of the preferred boundary conditions in
the $c=1$ moduli space, N stands for Neuman and D for Dirichlet boundary conditions \cite{Elitzur:1998va}}\label{c=1}
\end{figure}

One can associate a term in the boundary which measures the
effective number of degrees of freedom, this has been done by
various authors \cite{Affleck:1991tk}, \cite{Friedan:2003yc},\cite{Shatashvili:1993ps}.

 One can prove moreover  that one can define such a function whose value also decreases when the
theory flows on the boundary. All this while not requiring an
adjustment of the total central charge. What  happens for example
is that the theory flows from Neuman(N) to Dirichlet(D)  boundary
conditions \cite{Witten1992}, or in other words the branes may dissolve or may be
created under such a flow. In the figure below
(fig.\ref{c=1}) we give an example of a very simple
compactification for which one can identify what are the stable
configurations, describing when the system chooses to obey
Dirichlet and when the system chooses to obey  Neuman boundary
conditions \cite{Elitzur:1998va}.

\begin{figure}
\begin{center}
\includegraphics [scale=2, height=8cm]{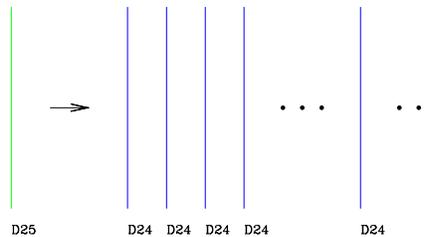}\\
\caption{A lattice of D24 branes is formed from a D25 brane in the presence of a boundary tachyon} \label{braneslattice}
\end{center}
\end{figure}

This can be used even further  if one changes the relevant
operator added on the boundary into a Sine-Gordon one. In that
case one can actually have situations where one breaks
translational invariance in space-time by a $D-25$ brane for
example dissolving into a lattice of $D-24$ branes
\cite{Harvey:2000na}, (fig.\ref{braneslattice}). Again such a
situation will lead also to a reduction of the original amount of
supersymmetry. Thus the idea of spontaneous breaking of spatial
translational invariance is demonstratively realized in String
Theory by the presence of open string tachyons.

\vspace{1 cm}

\section{Spontaneous Breaking of Time Translational Invariance and of Supersymmetry}

Next I will discuss a somewhat different mechanism  which may
allow  the possibility of a spontaneous breaking of
time-translational invariance. For that it is useful to consider
conformal and superconformal quantum mechanics. One way to
motivate the interest in such systems is to recall some basic
facts concerning the validity of a perturbative expansion.

Consider the Hamiltonian,

\begin{equation}
H= \frac{p_{q}^2}{2m} + \frac{1}{2} g q^n \ .
\end{equation}

 One may wonder  if it is possible to make a
meaningful perturbative expansion in terms of small
 or large $g$ or small or large $m$. To answer this question one needs to find out if
 one can remove the $g$,$m$ dependence from the operators, and relegate it to the total energy scale.
  This  type of rescaling  is used for discussing the harmonic oscillator.
One attempts  to define a new set of dimensionless canonical
variables
 $p_x, x$ that preserve the commutation relations
\begin{equation}
 [p_q,q]=[p_x,x] \hbar \ ,
 \end{equation}
 and
 \begin{equation}
   H=h(m,g) \frac{1}{2} (p_x^2 + x^n) \ .
 \end{equation}

   The following decomposition:

\begin{equation}
 q = f(m,g) x \comma \qquad p_q=  \frac{1}{f(m,g)}
p_x \comma
\end{equation}

gives

\begin{equation}
 2H = \frac{p_x^2}{m f^2(m,g)} + g f(m,g)^n x^n \;\; \comma
\end{equation}

 and
so one may choose
\begin{equation}
 g f(m,g)= \left(\frac{1}{ m f(m,g)^2}\right)^{ \frac{1}{n+2}}
\; \; .
\end{equation}

The Hamiltonian becomes:

\begin{equation}
 H= g^{1- \frac{n}{n+2}} m^{- \frac{n}{n+2}}\frac{1}{2} ( p_{q}^2 + q^n ) \; .
\end{equation}

The role of g and m is indeed just to determine the overall energy
scale. They may not serve as meaningful perturbation parameters.

 This does not apply to the special case of $n=-2$,  the case of conformal quantum mechanics, where  $g$  can be  a real
  perturbative  parameter.

\vspace{.3 cm}

\subsection{Conformal Quantum Mechanics: A Stable System With No Ground State Breaks Time Translational Invariance}

Consider  the Hamiltonian

\begin{equation}
 H= \frac{1}{2} (p^2 + g
x^{-2}) \label{cqm}
 \end{equation}

for a positive value of $g$  \cite{deAlfaro:1976je}.

  $H$ is part of the following algebra:

\begin{equation} [H,D] = iH
\comma [K,D] =iK \comma [H,K] =2 i D  \; \; .
\end{equation}

 It is  an SO(2,1) algebra, one representation of which is:

\begin{equation}
 D= - \frac{1}{4} (xp+px)  \comma K = \frac{1}{2} x^2
\end{equation}

 with $H$ is  given above. The Casimir is given by:

\begin{equation}
 \frac{1}{2} (HK +KH) - D^2 =  \frac{g}{4} -  \frac{3}{16} \ .
 \end{equation}
 In the Lagrangian formalism the system is described by:

\begin{equation}
 {\mathcal{L}}= \frac{1}{2} ( \dot{x}^2 - \frac{g}{x^2} ) \comma S =
\int \D t {\mathcal{L}} \ .
 \end{equation}

 Symmetries of the action $S$, and
not of  the Lagrangian ${\mathcal{L}}$ alone,  are given by:

\begin{equation}
t' =  \frac{at+b}{ct+d}  \comma x'(t')= \frac{1}{ct +d} x(t) \ ,
\end{equation}

  \begin{eqnarray}
   A= \left( \matrix{a &b \cr c&d\cr} \right) \comma \qquad \rm{det} A = ad-bc=1
\end{eqnarray}

 $H$ acts as  translation

 \begin{eqnarray}
 A_T=\left( \matrix{1&0\cr\udelta &1 } \right) \comma \qquad
 t'=t+\udelta \ .
\end{eqnarray}

 $D$ acts as dilation

 \begin{eqnarray}
  A_D= \left( \matrix{ \ualpha &0 \cr 0& {1 \over \ualpha}} \right) \comma \qquad
t'=\ualpha^2 t \ .
\end{eqnarray}

 $K$ acts as a special conformal
transformation
\begin{eqnarray}
A_K= \left( \matrix{1 & \udelta \cr 0&1} \right) \comma \qquad t'=
{t \over {\udelta t+1}} \ .
\end{eqnarray}

The spectrum of the Hamiltonian (\ref{cqm}) is the open set
$(0,\infty)$,
 the spectrum is therefore continuous and bounded from below. The wave  functions are given by:

\begin{equation}
 \psi_E(x)= \sqrt{x} J_{\sqrt{g + \frac{1}{4}} }(\sqrt{2E} x)
\; \; \quad E \neq 0 \; \; .
\end{equation}

The zero energy state is given by $\phi(x)=x^\ualpha$:

\begin{equation}
H= \left( -  \frac{d^2}{dx^2}  +  \frac{g}{x^2}  \right) x^\ualpha
=0 \; \; .
\end{equation}

This implies

\begin{equation}
 g= -\ualpha(\ualpha-1) \ ,
 \end{equation}

 solving this equation
gives
\begin{equation}
 \ualpha=- \frac{1}{2} \pm  \frac{\sqrt{1+4g}}{2} \ .
 \end{equation}

This gives rise to two independent solutions and by completeness
these are all the solutions. The case  $\ualpha_+ >0$,  does not
lead to a normalizable solution since the function diverges at
infinity. $\ualpha_- <0$, is not normalizable either since the
function diverges at the origin (a result of the scale symmetry).

Thus, there is no normalizable (not even plane wave normalizable)
E=0 solution!

\begin{figure}[hbtp]
\centering
\includegraphics[width=8.0 truecm]{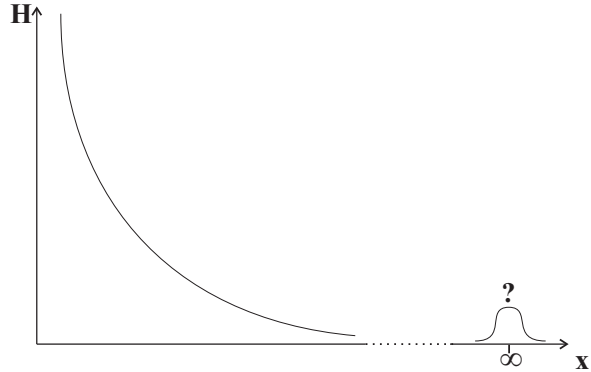}
\caption[]{Their  no normalisable ground state for this
potential} \label{figure}
\end{figure}

Most of the analysis in field theory proceeds by identifying a
ground state and the fluctuations around it. How do we deal with a
system in the absence of a ground state?

One possibility is to accept this as a fact of life. Perhaps it is
possible to view this as similar to cosmological models that also
lack a ground state, such those with Quintessence. In field theory
such systems have no finite energy states in the spectrum at all.
Only time dependent states are allowed. In the presence of an
appropriate cutoff and in quantum mechanics it is only the
potential lowest energy state which is disallowed.

Another possibility is to define a new evolution operator that
does have a ground state

\begin{equation}
 G= u H +v D +w K \ .
\end{equation}

 This operator has a ground state if $v^2-4uw <0$.
Any choice explicitly breaks  scale invariance. Take for example

\begin{equation}
G= \frac{1}{2} \left( \frac{1}{a} K + a H \right) \equiv R \ ,
\label{Rop}
\end{equation}

 $a$
has the dimension of a length. The eigenvalues of $R$ are

\begin{equation}
r_n = r_0+ n \comma \qquad r_0= \frac{1}{2} \left(1+ \sqrt{g+
\frac{1}{4} }\right) \ .
\end{equation}

This is a breaking of scale invariance by a dictum and not by the
dynamics of the system. Nevertheless  it is very interesting to
search for a physical interpretation of this. Surprisingly this
question arises in the context of black hole physics \cite{Claus:1998ts}. Consider a
particle of mass $m$ and charge $q$ falling into a charged black
hole.
 The black hole is BPS, meaning that its mass  $M$ and charge  $Q$ are related, in the appropriate unites,  by $M=Q$.

  The blackhole metric and vector potential are given by:

\begin{equation}
 \D s^2=- \left(1+ \frac{M}{r}\right)^{-2} \D t^2 + \left( 1+ \frac{M}{r} \right)^2 (\D r^2 + r^2
 \D  \Omega^2) \comma \quad A_t=  \frac{r}{M}
\end{equation}

Now consider the near Horizon limit, i.e. $r<<M$, which we will
 reach  by taking  $M \rightarrow \infty$ and keeping $r$ fixed.
This produces an $AdS_2 \times S^2$ geometry
 \begin{equation}
 \D s^2=- \left( \frac{r}{M} \right)^2 \D t^2 + \left( \frac{M}{r} \right)^2 \D r^2 + M^2 \D
 \Omega^2 \ .
\end{equation}

We also wish to keep  $M^2 (m-q)$ fixed as we scale $M$. This
means we must scale
 $(m-q) \rightarrow 0$, that is the particle itself becomes BPS in the limit.

The Hamiltonian for this  falling in particle in this limit is
given by our old friend:

\begin{equation}
H=  \frac{p_{r}^2}{2m} + \frac{g}{2r^{2}} \comma \quad g =
8M^2(m-q) +  \frac{4l(l+1)}{M} \ .
 \end{equation}

  For $l = 0$, we have $g > 0$ and there is no ground
state. This is associated with the coordinate
 singularity at the Horizon. The change in evolution operator is now
associated with a change of time coordinate. One for which the
world line of a static particle passes
 through the black hole horizon instead of remaining in the exterior of the space time.
 In any case, the consequence of removing the potential lowest energy state
  of the system from the spectrum can be described as a breaking of time translational invariance.

\subsection{Superconformal Quantum Mechanics:A Stable System With No Ground State Breaks Also Supersymmetry}

The bosonic conformal mechanical system had no ground state. The
absence of a $E=0$ ground state in the supersymmetric context
leads
 to the breaking of supersymmetry. This breaking has a different flavor from that
 which was discussed for the spatial translations.
 We next examine the supersymmetric version of conformal quantum
 mechanics \cite{Fubini:1984hf}, \cite{Akulov:uh},
 to see if indeed supersymmetry is  broken.
 The superpotential is chosen to be

 \begin{equation}
W(x)=\frac{1}{2} g\; \log x^2 \comma
\end{equation}
 yielding the Hamiltonian:

\begin{equation}
 H= \frac{1}{2} \left[ \left( p^2 + \left( {{ \D w} \over
{ \D x}}\right)^2 \right) 1 -   \frac{d^2W}{dx^2}\sigma_3  \right]
\; \; .
\end{equation}

Representing $\psi$ by $\frac{1}{2}  \sigma_-$  and $\psi^*$by
$\frac{1}{2} \sigma_+$ gives the supercharges:

\begin{equation}
Q = \psi^+ \left(- \I p +  \frac{\D W }{dx} \right) \comma Q^+
=\psi \left( \I p +
 \frac{ \D w }{ \D x} \right) \ .
\end{equation}

 One now has a larger algebra, the superconformal
algebra,

 \begin{eqnarray}
\{ Q,Q^+ \} &=& 2H  \comma \{ Q, S^+ \} =g-B +2iD  \ , \nonumber \\
\{ S,S^+ \} &=&2K   \comma \{ Q^+,S \}=g-B-2iD \ .
 \end{eqnarray}

  A realization is:
  \begin{equation}
   B=\sigma_3 \comma S=\psi^+ x
\comma S^+ = \psi x \; \; .
 \end{equation}

The zero energy solutions are
\begin{equation}
exp(\pm W(x) ) =
x^{\pm g} \comma
\end{equation} neither solution is normalizable.

$H$ factorizes:

\begin{eqnarray}
 2H= \left( \matrix{ p^2 + \frac{g(g+1)}{x^2}  & 0 \cr 0 & p^2 + \frac{g(g-1)}{x^2}}
 \right) \ ,
\end{eqnarray}

 and we may solve for the full spectrum:
\begin{equation}
\psi_E(x)= x^{1/2} J_{\sqrt{\unu} }(x\sqrt{2E} )\comma \qquad E
\neq 0 \comma
\end{equation}

 where $\unu=g(g-1)+ 1/4$ for $N_F=0$ and
$\unu=g(g+1)+ 1/4$ for $N_F=1$.

 The spectrum is continuous and  there is no normalizable zero energy state.
One must interpret the absence of a normalizable ground state. It
is also possible to  define a new operator which has a
normalizable ground state.
 By inspection the operator (\ref{Rop}) can be used, provided one makes the following identifications:

\begin{eqnarray}
N_F&=&1 \qquad g_B =g_{susy}(g_{susy}+1)\comma  \nonumber \\
N_F&=&0 \qquad g_B=g_{susy}(g_{susy}-1)\ .
\end{eqnarray}

Thus the spectrum differs between the $N_F=1$ and $N_F=0$ sectors
and
 supersymmetry would be broken. One needs to define a whole new set of operators:

\begin{eqnarray}
M=Q-S \qquad M^+=Q^+-S^+  \nonumber \\
N=Q^+ +S^+ \qquad N^+=Q+S^+
\end{eqnarray}

 which produces the
algebra:
\begin{eqnarray}
\frac{1}{4} \{ M, M^+ \}&=& R + \frac{1}{2} B - \frac{1}{2}  g \equiv T_1  \ , \nonumber  \\
 \frac{1}{4} \{ N, N^+ \}&=& R + \frac{1}{2} B + \frac{1}{2}  g \equiv T_2  \ , \nonumber    \\
\frac{1}{4} \{ M, N \} &=& L_- \qquad  \frac{1}{4} \{ M^+,N^+ \} = L_+   \ , \nonumber  \\
L_\pm &=& - \frac{1}{2} (H- K \mp 2 i D)
\end{eqnarray}

$T_1, T_2, H$ have a doublet spectra. ``Ground states'' are given
by:

\begin{equation}
 T_1|0>=0 \; ; \quad T_2 |0>=0 \; ; \quad  H|0>=0
\; \; .
\end{equation}

In this setup one can also exhibit  \cite{Fubini:1984hf} how the
in the presence of a breaking of a space time symmetry, global
$\mathcal{N}=2$  can be broken only to $\mathcal{N}=1$.  A
physical context arises when one considers a supersymmetric
particle falling into a black hole \cite{Claus:1998ts},
\cite{Kallosh:1999mi}. This is the supersymmetric analogue of the
situation already discussed.

One should  mention again that there is a dictum in the way one
has broken scale/conformal invariance in the problem. It is
amusing to mention that if one takes the dictated ground state and
decomposes it in terms of the energy eigenstates, then one usually
gets that the new ground state looks like a thermal distribution
of the old ground states. This looks very attractive and it is
related to black holes, which as mentioned above do come up.

Another example where such breakdown of time-translational
invariance may occur  is  the Liouville model.  Also there, there
is no normalizable ground state. For works on the possible
breakdown of  translational invariance in the two dimensional
Liouville model, see \cite{D'Hoker:1983ef},\cite{Bernard:1983ip}.

 Beyond $d = 2$, we can also mention that in four
 dimensions in $\mathcal{N} = 1$ supersymmetric theories,
 where the number of flavors   $N_{F}$ is smaller then the
 number of colors $  0 < N_{F} < N_{C}$, one gets \cite{Affleck:1983mk},\cite{Taylor:1982bp}. This is another  situation
 where the spectrum is bounded from below but there is no ground
 state (fig.\ref{breakings}). The spectrum is open, and actually in the presence of a cutoff such systems
 have no finite energy states at all, which is a very interesting
 as far as Cosmology is concerned.

\section{Spontaneous Breaking of Conformal Invariance}

Fubini also suggested to discuss such situations in a general
number of dimensions \cite{Fubini}. He researched it in a
scientific environment which did  not yet   fully realize  that
interacting finite theories might exist in various number of
dimensions. Therefore much of his analysis was of a classical
nature. He emphasized the conformal features of the system, and we
are going to discuss the breakdown of conformal invariance. The
discussion of the breakdown of  time translational invariance
brought us to conformal theories and now we  are discussing  also
the breakdown of the conformal invariance.

 If one considers a theory with only one
scalar field, a general classic conformal invariant is given by
the following Lagrangian

\begin{equation}
\mathcal{L} = \frac{1}{2}\partial_{\umu}\phi \partial^{\umu}
\phi - g \phi^{\frac{2d}{d-2}}.
\end{equation}

 The symmetry of the system is  the bosonic,  $O(d,2)$ symmetry, and
 the generator are $M_{\umu\unu}$, $P_{\umu}$, of the Poincar\`e group, the special conformal transformation generator  $K_{\umu}$,
  and the dilatation $D$. The dictum of Fubini in this case is
  that the ground state is not translational invariant, this is
  not accompanied by any dynamical calculation.   The vacuum expectation value $< \phi(x)>$,
is $x$ dependent, and actually it looks very much like an
instanton

\begin{equation}
 < \phi(x)> = b \left( \frac{a^{2} +
x^{2}}{2a}\right)^{-\frac{d-2}{2}} \ ,
 \end{equation}

 which is a solution of the equation of motion

\begin{equation}
\partial^{2} \phi(x) - 2g \frac{d}{d-2}\phi^{\frac{d+2}{d-2}}(x) = 0 \ .
\end{equation}

\begin{figure}
\begin{center}
\includegraphics [scale=2, height=7cm]{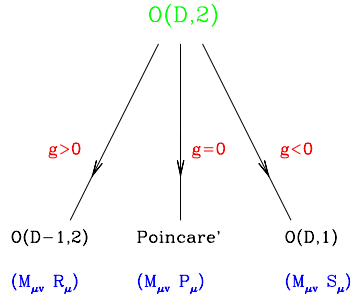}\\
\caption{The sign of the quartic coupling  $g$ determines the
symmetry breaking patterns  of the symmetry group $O(d,2)$}
\label{breakings}
\end{center}
\end{figure}

By choosing this to be the vacuum, (again I emphasize, this is by
dictum), one breaks down the $O(d,2)$ symmetry, (as in fig.
\ref{breakings}) in the following fashion: if the coupling $g$ of
the scalar self-interaction is positive the theory breaks down to
$O(d-1,2)$ and the resulting symmetries are $M_{\umu \unu}$,
$R_{\umu}$. If $g<0$  the symmetry breaks to $O(d-1)$, generated
by $M_{\umu\unu}$, $S_{\umu}$, where

\begin{equation}
 S_{\umu} = \frac{1}{2} \left( a P_{\umu} - \frac{1}{a}
K_{\umu} \right) \ .
\end{equation}

If $g  =0$ one remains with Poincar\'e invariance, (fig.
\ref{breakings}). In the de Sitter example, which occurs for
$g>0$, one can show again that there are signatures of
temperature. A question which at the time seemed interesting was:
Does a spontaneous breaking of conformal invariance  require also
the breakdown of translational invariance? Examples were since
found where this is not the case. Counter examples to the  idea
that the breaking of conformal invariance must drive a breaking of
Supersymmetry were discovered, We will discuss in more detail some
such examples. One can break scale invariance without breaking
rotational or translational invariance. We also mention briefly
that conformal invariance and scale invariance are not always
equivalent, and in a set of works,(see for example
\cite{Polchinski:1987dy}), one can show that scale invariance
leads under some certain conditions to conformal invariance.

 That is the  case when  the spectrum of the theory is discrete, such as
for a  two dimensional sigma model description in which the target
space is compact. But for non-compact target spaces one can find
counter examples \cite{Elitzur:1994ri}, in which scale invariance does not lead to conformal
invariance.
 In recent years it has been fully realized that theory which are
quantum mechanically scalar invariant and finite may exist in  $d
= 2,3,4,5,6$ dimensions. Such theories can exhibit spontaneous
breaking, for example the $d=4$, $\mathcal{N} = 4$ Super Yang-Mils
with $SU(N)$ gauge group is characterized  by the following
spectrum

$$(A^{a}_{\umu}, \lambda^{a}, \phi^{a} + i\varrho^{a}).$$

The theory is parameterized by the complex parameter $ig +
\theta$, where $g$ is the coupling constant and $\theta$
 is the $\theta$ angle . Such a theory has flat directions
 which allow   phases where either $<\phi>$
vanishes and the theory is realized in a conformal manner, or a
phase in which $<\phi> \ne 0$ along flat directions. This is the
Coulomb phase,
 in which the gauge group $SU(N)$ may be reduced all the way to
 $U(1)^{N}$, where $N$ is the rank of the gauge group.
 This is the maximum possible breaking of the gauge group
  when the fields are in the adjoint representation. In such a
  case, scale invariance is broken spontaneously and the
  vacuum energy remains zero, and there is no breakdown of either
  translational invariance or Supersymmetry. Such a theory will have a Goldstone
  boson, associated with the spontaneous breaking of scale
  invariance, which is called the dilaton.
  This is a \emph{true} dilaton worthy of his name.
  It is interesting to note that in such a system
 the vacuum energy is not influenced by  the value of $<\phi>$,
  and it vanishes in all the  phases.

\vspace{.5 cm}
 \section{ $O(N)$ Vector Models in $d=3$: Spontaneous Breaking of Scale Invariance and The Vacuum Energy}

The next example that we have is related to the spontaneous
breaking of scale invariance in a three dimensional bosonic
theory. Such a theory describes the mixing of $He_{3}$ and
$He_{4}$, (see \cite{Amit:1984ri} and references there in).

The most general Lagrangian describing such a system is

\begin{equation} \mathcal{L} = \frac{1}{2}\partial_{\umu}\vec{\phi} \
\partial^{\umu}\vec{\phi} - \frac{1}{2}\lambda_{2}(\vec{\phi})^{2}
+ \frac{\lambda_{4}}{4N}(\vec{\phi})^{4} +
\frac{\lambda_{6}}{N^{2}}(\vec{\phi})^{6}, \end{equation}

and it can be treated at $d = 3 - \varepsilon$. The system  has
two order parameters $<(\vec{\phi})^{2}>$, and $<\vec{\phi}>$.

\begin{figure}
\begin{center}
\includegraphics [scale=2, height=7cm]{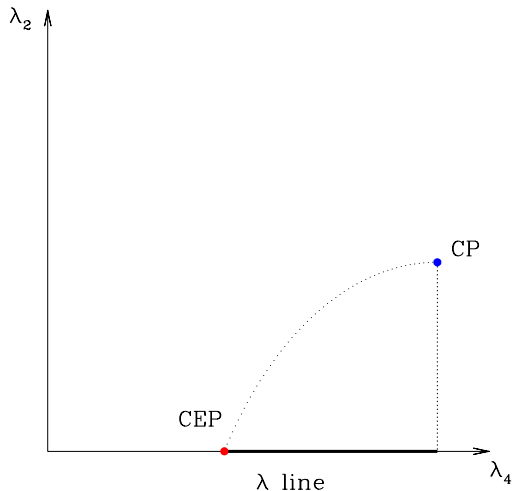}\\
\caption{The Phase Diagram of a $d=3-\varepsilon$ Conformal Theory, in three dimensions the $CP$ and $CEP$ points coincide to produce
 a flat direction, see \cite{Amit:1984ri} for more details}  \label{He}
\end{center}
\end{figure}

In a classical analysis performed for $d = 3 - \varepsilon$, when
the sign of $\lambda_{2}$ changes then $<\vec{\phi}>$ is produced.
However, $<(\vec{\phi})^{2}> \ne 0$ even for $\lambda_{2} > 0$,
which is exemplified by the diagram shown in fig. \ref{He}.

When one goes to three dimensions, the point  which is denoted by
CP, which is a critical point, and the point CEP which is the
critical end point, do actually meet together and lead to a very
interesting structure. Going directly to $d=3$, one can write down
the $O(N)$ vector model written below

\begin{equation} \mathcal{L} = \frac{1}{2}\partial_{\umu}\vec{\phi} \
\partial^{\umu}\vec{\phi} - \frac{1}{2}\lambda_{2}(\vec{\phi})^{2}
+ \frac{\lambda_{4}}{4!N}(\vec{\phi})^{4} +
\frac{\lambda_{6}}{6!N^{2}}(\vec{\phi})^{6} \ . \end{equation}

It should be emphasized  that everything said depends on the very
specific manner of taking the limit. One first keeps the cutoff
 $\Lambda$ fixed and takes $N \rightarrow \infty$, by performing a
 functional integral or selecting a subset of diagrams, and only then does one remove the cutoff,
 sending it to infinity, setting the renormalized quadratic and
 quartic  couplings  to zero. Such a system turns out to be not
 only classically conformally invariant, but also quantum
 mechanically, having a vanishing beta function \cite{bmb}. We next elaborate on such systems.



Let us now review some more known facts about the three dimensional
theory once a classically marginal operator,$(\vec{ \phi }^2)^3$,
is added~\cite{bmb}. For any finite value of $N$, the coupling
$g_6$ of this operator  is infrared free quantum mechanically, as
the marginal operator gets a positive anomalous dimension already
at one loop. This implies that the theory is only well defined for
zero value of the coupling of this operator. In the presence of a
cutoff interacting particles have mass of the order of the cutoff.
At its tri-critical point the $O(N)$ model in three dimensions is
described by the Lagrangian

 \begin{equation}
 \mathcal{L} = \frac{1}{2} \partial_{\umu} {\vec{ \phi}} \ \partial^{\umu}
{\vec{ \phi}} + \frac{1}{6N^2}g_6 (\vec{ \phi}^2)^3~,
\end{equation}

 where the fields $\vec{ \phi}$ are in the vector
representation of $O(N)$.

In the limit $N \rightarrow \infty$ \cite{bmb}

 \begin{equation}
  \ubeta_{g_6}=0  \ .
  \end{equation}

$1/N$ corrections break conformality. In the large $N$ limit,
then, $g_6$ is a modulus.  It turns out that there is no
spontaneous breaking of the $O(N)$ symmetry and it is instructive
to write the effective potential in terms of an $O(N)$ invariant
field,

\begin{equation}
 \sigma = \vec{ \phi}^2 \ .
 \end{equation}

The effective potential is \cite{Amit:1984ri}:

 \begin{equation}
  V(\sigma)= f(g_6)|\sigma|^3  \; \; ,
 \end{equation}
  where:

   \begin{equation}
   f(g_6)= g_c-g_6
 \end{equation}

  with
  \begin{equation}
 g_c=(4\upi)^2~.
   \end{equation}

   The system has various phases. For values of $g_6$ smaller than $g_c$, i.e. when
$f(g_6)$ is positive,  the system consists of $N$ massless
non-interacting $\phi$  particles. These particles do not interact
in the infinite $N$ limit; thus, correlation functions do not
depend on  $g_6$. For the special value $g_6=g_c$, $f(g_6)$
vanishes and a flat direction in $\sigma$ opens up: the
expectation value of $\sigma$ becomes a modulus. For a zero value
of  this expectation value, the theory continues to consist of $N$
massless $\phi$ fields. For any non-zero value of the expectation
value the system has $N$
 massive $\phi$ particles.
 All have the same mass due to the unbroken $O(N)$ symmetry. Scale invariance is
broken spontaneously though the vacuum energy still vanishes. The
Goldstone boson associated with the spontaneous breaking of scale
invariance, the dilaton, is massless and identified as the $O(N)$
singlet field $\udelta\sigma\equiv \sigma-\langle \sigma \rangle$.
All the particles are non-interacting in the infinite $N$ limit.
This theory is not conformal: in the infrared limit, it flows to
another theory containing a single, massless, $O(N)$-singlet
particle.
  For larger values of $g_6$ the exact potential is unbounded from below. The
system is unstable (in the supersymmetric case the potential is
bounded from below and the larger $g_6$ structure is similar to
the smaller $g_6$ structure~\cite{bhm}). This case is useful to illustrate the fate of some
gravitational instabilities \cite{Elitzur:2005kz}.
 Actually this instability
is an artifact of the dimensional regularization used above, which
does not respect the positivity of the renormalized field
$\sigma$.
In any case a more careful analysis~\cite{bmb} shows that  the
apparent instability reflects the inability to define a
renormalizable interacting theory.  All the  masses are of the
order of the cutoff, and there is no mechanism to scale them down
to low mass values. In other words, the theory depends strongly on
its UV completion.

 This is summarized in Table 1.

\begin{table}[t]\centering
\caption{Marginal Perturbations of the $O(N)$ Model}
\begin{tabular}{ccccc}
$f(g_6)$   & $|\langle\sigma\rangle|$ & S.B. & masses & $V$  \\
$f(g_6) > 0$  &   0     &      No & 0  &0 \\
$f(g_6)=0$   &    0 & No &0&0 \\
$f(g_6)=0$   & $\neq 0$& Yes & Massless dilaton, $N$ & 0\\
&&& particles of equal mass & \\
$f(g_6)<0$   &  $\infty$   & Yes,  & Tachyons or masses &
$-\infty$
\\ &&but ill defined & of order the cutoff  &
\end{tabular}
\end{table}

There, S.B. denotes spontaneous symmetry breaking of scale
invariance and $V$ is the vacuum energy. For $f(g_6) <0$ the
theory is unstable. Note that the vacuum energy  always vanishes
whenever the theory is well-defined.

When  $\langle\sigma\rangle \neq 0$ and the scale invariance is
spontaneously broken, one can write down the effective theory for
energy scales below $\langle\sigma\rangle$, and integrate out the
degrees of freedom above that scale. The vacuum energy remains
zero however, and it is not proportional to
$\langle\sigma\rangle^3$,\cite{Amit:1984ri},\cite{Einhorn:1985wp},\cite{Rabinovici:1987tf},\cite{Rabinovici:1987tf},
\cite{fub},\cite{br}, as might be expected naively.

\vspace{.5 cm}

 For completeness we note that by adding more vector
fields one has also phases in which the internal global $O(N)$
symmetry is spontaneously broken.

An  example is  the $O(N) \times O(N)$ model
\cite{Rabinovici:1987tf} with two fields in the vector
representation of $O(N)$, with Lagrangian:

\begin{eqnarray}
 \mathcal{L} &=& \partial_{\umu} {\vec{\phi}_1} \cdot \partial^{\umu}
{\vec{\phi}_1} +\partial_{\umu} {\vec{\phi}_2} \cdot
\partial^{\umu} {\vec{\phi}_2} +
\lambda_{6,0} (\vec{\phi}_1^2)^3 +  \nonumber  \\
&& \lambda_{4,2} (\vec{\phi}_1^2)^2 (\vec{\phi}_2^2) +
\lambda_{2,4} (\vec{\phi}_1^2) (\vec{\phi}_2^2)^2 + \lambda_{0,6}
(\vec{\phi}_2^2)^3 \ .
\end{eqnarray}

 Again, the $\ubeta$ functions
vanish in the strict $N \rightarrow \infty $ limit. There are now
two possible scales, one associated with the breakdown of a global
symmetry and another with the breakdown of scale invariance. The
possibilities are summarized by the table below:

\begin{eqnarray}
\matrix{ & O(N)&O(N)& {\rm{scale}}& {\rm{massless}}&{\rm{massive}}
&V \cr
         & +&+&+&{\rm{all}}&{\rm{none}}&0 \cr
         &-&+&-&(N-1)\upi's, D&N,\sigma&0 \cr
&+&-&-&(N-1)\upi's, D & N,\sigma&0 \cr &-&-&-& 2(N-1)\upi's,D &
\sigma& 0 \cr }
\end{eqnarray} \label{e8}

 Again in all cases the vacuum energy
vanishes. Assume a hierarchy of scales where the scale invariance
is broken at a scale much above the scale at which the $O(N)$
symmetries are broken. One would have argued that one would have
had a low energy effective Lagrangian for the massless  pions and
dilaton with a vacuum energy given by the scale at which the
global symmetry is broken. This is not true, the vacuum energy
remains zero. This system has a critical surface, on one patch the
deep infrared theory contains only one massless particle: an
$O(N)\times O(N)$ singlet. For the other patches the deep infrared
theory is described by $O(N)$ massless particles, most of which
are not $O(N)$ singlets. The dimension of the surface for which spontaneous
symmetry breakings occur is smaller then that of the full space of parameters.
I will not consider spontaneous symmetry breaking as fine tuning.
 \vspace{1 cm}

In general, effective  field theories should have all possible
symmetries of the underlying theory, whether they are  realized linearly or non-linearly.
In finite scale invariant theories the vacuum energy $E_{vac}$
should be determined by all scales and symmetries involved. It should
have the same value, (zero in this case), in all phases of the
system whether or expectation values are formed. This punches a
hole in Zaldowitch  like arguments \cite{zel}, and offers a
different view on the gravity of the Cosmological Constant problem
\cite{Weinberg:1988cp}.
 If the theory
has a global scale invariance, which is spontaneously broken, it
will produce a dilaton. The question is: Where is the dilaton? The
dilaton should be a massless field. Several authors  ,
\cite{Damour:1994zq},\cite{Damour:2002nv} tried to check the
possibilities that the dilatons might exist, noting that the
dilaton must be a massless Goldstone boson. Under certain
assumptions, one finds out that actually in certain models having
a massless dilaton would not violate experimental data. Perhaps it
even predicts deviations of the equivalence principle from Galileo
famous experiment just below the
present experimental sensitivity $\delta a / a \sim 10^{-12}$.


 This is done under the assumption that the dilaton couples
in the following universal fashion

\begin{equation} \mathcal{L} = F(\Phi ) \left( R - F^{2} + 2[ \nabla^{2} \Phi
- (\nabla \Phi )^{2} ] \right) \ .  \end{equation}

It could also happen that the dilaton gets swallowed in some Higgs
like mechanism. One should also mention that if kinematically a
finite scale-invariant is forced by some super-selection rule
 (such as having a non trivial monopole number \cite{rabidea}) into a certain
solitonic sector, then the rest energy of the system should be
accounted for and the vacuum energy will be slightly lifted from
zero. Let us finish this section by noticing an amusing thing,
there are various  solutions that go under the name of Randall and
Sundrum  \cite{Randall:1999vf}. One of the constructions contains  two types of branes,
near the boundary of the space there is a Planck brane with
tension $T_{1}$, which is fine tuned so to have zero Cosmological
Constant. Then at a certain distance, very deep inside the bulk
theory, one places the TeV brane, it has negative tension and the
tension is again fine tuned, so that the Cosmological Constant
vanishes also on that brane.

\begin{figure}
\begin{center}
\includegraphics[scale=2,height=9cm]{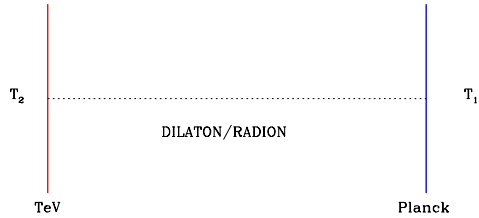}\\
\caption{Planck brane and Tev brane} \label{RZ}
\end{center}
\end{figure}

The two branes are separated by some distance which in
\cite{Rattazzi:2000hs}
 is associated to massless particle, which
is the dilaton or the radion, (see fig. \ref{RZ}).

 In principle, there are circumstances
where this distance is not fixed, and there are several possible
situations whose outcome is very similar to that one discussed in
the $d=3$ conformal theory. If the sum of the tensions  $T_{1} +
T_{2}$ is arranged to vanish, then the system behaves as a
spontaneously broken system,  the magnitude of the vev of the
field is the distance between the two branes.

If $T_{1} + T_{2} > 0$ the two branes actually are attracted to
sit one on top of the other, and when $T_{1} + T_{2} < 0$  the
branes repel , the system is unstable and as a result one of the
branes is exiled to infinity.

\vspace{.5 cm}

These three examples are in full correspondence with the
conditions on the coefficients of the $(\vec{\phi})^{6}$ theory
that we discussed above. The difference between the two theories,
and an important difference is that in case of the
$(\vec{\phi})^{6}$ theory we are certain that  in the large $N$
limit, the theory is indeed finite quantum mechanically.

For the case of $( \vec{\phi})^{4}$ we don't have such an
assurance, and it would be nice to find a system for which we are
guaranteed to be finite also quantum mechanically, which exhibits
the same type of behavior.

\vspace{.7 cm}


 \vspace{.7 cm}

\subsection{Conclusions}
\noindent

$\bullet$ Spontaneous breaking of translational and rotational
symmetry are possible. It fits  data  for many phases of matter,
and it may have a manifestation in the dynamics of
compactification.

\vspace{.5 cm}

$\bullet$ Conformal/Scale invariant theories which are stable but
have no ground states indicate a new mechanism of breaking time
translational invariance as well as supersymmetry.

\vspace{.5 cm}

 $\bullet$  A finite scale invariant theory has the same (vanishing) vacuum
energy in all its phases.

\vspace{.5 cm}

  $\bullet$    It is a great privilege to recognize Gabriele's outstanding
contributions.

 \vspace{3 cm}

\section*{Acknowledgements}
The author thanks Matteo Cardella for various discussions on this
manuscript. The author wishes to thank his various collaborators
on these subjects, especially W. Bardeen, S. Elitzur, M. Einhorn,
A. Forge, A. Giveon, M. Porrati, A. Schwimmer, AND G. Veneziano.


\vspace{1 cm}


\begin{thebibliography}{99.}


\bibitem{Fubini:1984hf}
S.~Fubini and E.~Rabinovici,
Nucl.\ Phys.\ B  \textbf{ 245} (1984) 17



\bibitem{Hughes:1986dn}
  J.~Hughes and J.~Polchinski,
  Nucl.\ Phys.\  B  \textbf{278} 147 (1986).






\bibitem{Landau}
 L.D.Landau, Phys. Z. Soviet II \textbf{26}, 545 (1937)


\bibitem{EFR}
  S.~Elitzur, A.~Forge and E.~Rabinovici,
  ``Some global aspects of string compactifications,''
  Nucl.\ Phys.\ B \textbf{ 359} (1991) 581.



\bibitem{Baym}
 C. Baym, H.A. Bethe, and C.J. Pethick, Nucl Phys A \textbf{175},
 25 (1971)










\bibitem{Alexander}
 S. Alexander: Symmetries and Broken Symmetries in Condensed
 Matter Physics, ed. N. Boccara (IDSET Paris,1981) p.141, and references therein.



\bibitem{Zamolodchikov}
 A.B. Zamolodchikov, Sov. Phys. JETP Lett. \textbf{43}, 730 (1986)


\bibitem{Ludwig}
A.W.W. Ludwing and J.L. Cardy, Nucl. Phys. B\textbf{285} [FS19],
687 (1987)



\bibitem{Elitzur:1998va}
  S.~Elitzur, E.~Rabinovici and G.~Sarkissian,
  Nucl.\ Phys.\  B  \textbf{541}, 246 (1999)


\bibitem{Affleck:1991tk}
  I.~Affleck and A.~W.~W.~Ludwig,
  Phys.\ Rev.\ Lett.\  \textbf{ 67}, 161 (1991)




\bibitem{Shatashvili:1993ps}
  S.~L.~Shatashvili,
  Alg.\ Anal.\  {\bf 6}, 215 (1994)





\bibitem{Friedan:2003yc}
  D.~Friedan and A.~Konechny,
  Phys.\ Rev.\ Lett.\  {\bf 93}, 030402 (2004)
  [arXiv:hep-th/0312197].




\bibitem{Witten1992}
  E.~Witten,
  Phys.\ Rev.\  D {\bf 47}, 3405 (1993)











\bibitem{Harvey:2000na}
  J.~A.~Harvey, D.~Kutasov and E.~J.~Martinec,
  ``On the relevance of tachyons,''
  arXiv:hep-th/0003101.


\bibitem{deAlfaro:1976je}
V.~de Alfaro, S.~Fubini and G.~Furlan,
Nuovo Cim.\ A  \textbf{ 34} (1976) 569


\bibitem{Akulov:uh}
V.~P.~Akulov and A.~I.~Pashnev,
Theor.\ Math.\ Phys.\  \textbf{ 56} (1983) 862 [Teor.\ Mat.\ Fiz.\
\textbf{ 56} (1983) 344]






\bibitem{Claus:1998ts}
P.~Claus, M.~Derix, R.~Kallosh, J.~Kumar, P.~K.~Townsend and
A.~Van Proeyen, 
Phys.\ Rev.\ Lett.\  \textbf{ 81} (1998) 4553



\bibitem{Kallosh:1999mi}
R.~Kallosh, ``Black holes and quantum mechanics,''
arXiv:hep-th/9902007.

















\bibitem{D'Hoker:1983ef}
  E.~D'Hoker and R.~Jackiw,
  Phys.\ Rev.\ Lett.\  \textbf {50}, 1719 (1983)




\bibitem{Bernard:1983ip}
  C.~W.~Bernard, B.~Lautrup and E.~Rabinovici,
  Phys.\ Lett.\  B \textbf{134}, 335 (1984)





\bibitem{Taylor:1982bp}
  T.~R.~Taylor, G.~Veneziano and S.~Yankielowicz,
  Nucl.\ Phys.\  B {\bf 218}, 493 (1983)


\bibitem{Affleck:1983mk}
I.~Affleck, M.~Dine and N.~Seiberg,
  Nucl.\ Phys.\  B  \textbf{241}, 493 (1984).



\bibitem{Fubini}
Sergio Fubini,
 ``A New Approach To Conformal Invariant Field Theories,''
 CERN-TH-2129, Feb 1976. 46pp.
  Published in Nuovo Cim. \textbf{A34}: 521,1976.







\bibitem{Polchinski:1987dy}
  J.~Polchinski,
  Nucl.\ Phys.\   \textbf{303}, 226 (1988)



\bibitem{Elitzur:1994ri}
  S.~Elitzur, A.~Giveon, E.~Rabinovici, A.~Schwimmer and G.~Veneziano,
  Nucl.\ Phys.\  B  \textbf{435}, 147 (1995)




\bibitem{Amit:1984ri}
  D.~J.~Amit and E.~Rabinovici,
 Nucl.  Phys.   B \textbf{257}, 371 (1985)

\bibitem{bmb}W.~A.~Bardeen, M.~Moshe and M.~Bander,
Phys.\ Rev.\ Lett.\  \textbf{ 52}, 1188 (1984).



\bibitem{bhm}
W.~A.~Bardeen, K.~Higashijima and M.~Moshe,
  Nucl.\ Phys.\ B \textbf{250}, 437 (1985)



\bibitem{Elitzur:2005kz}
  S.~Elitzur, A.~Giveon, M.~Porrati and E.~Rabinovici,
  JHEP {\bf 0602}, 006 (2006)






\bibitem{br}
D.~S.~Berman and E.~Rabinovici, ``Supersymmetric gauge theories,''
arXiv:hep-th/0210044.

\bibitem{Einhorn:1985wp}
  M.~B.~Einhorn, G.~Goldberg and E.~Rabinovici,
  Nucl.\ Phys.\  B  \textbf{256}, 499 (1985).


\bibitem{Rabinovici:1987tf}
  E.~Rabinovici, B.~Saering and W.~A.~Bardeen,
  Phys.\ Rev.\  D  \textbf{36}, 562 (1987).



\bibitem{zel}
Ya. B. Zeldovich, Sov. Phys. Uspekhi \textbf{11} (1968) 381.


\bibitem{Weinberg:1988cp} S.~Weinberg, 
 Rev.\ Mod.\ Phys.\  \textbf{61} (1989)

\bibitem{Damour:1994zq}
  T.~Damour and A.~M.~Polyakov,
  Nucl.\ Phys.\  B  \textbf{423}, 532 (1994)






\bibitem{Damour:2002nv}
  T.~Damour, F.~Piazza and G.~Veneziano,
  Phys.\ Rev.\  D  \textbf{ 66}, 046007 (2002)




\bibitem{Randall:1999vf}
  L.~Randall and R.~Sundrum,
  Phys.\ Rev.\ Lett.\  {\bf 83}, 4690 (1999)





\bibitem{Rattazzi:2000hs}
  R.~Rattazzi and A.~Zaffaroni,
  JHEP   \textbf{0104}, 021 (2001)







\bibitem{fub}
 E. Rabinovici in  W.~M.~Alberico and S.~Sciuto: Symmetry And Simplicity In
Physics. In: \textit{Proceedings, Symposium On The Occasion Of
Sergio Fubini's 65th Birthday}, Turin, Italy, February 24-26,
1994, Singapore (World Scientific 1994) p 220.

\bibitem{rabidea}
E. Rabinovici, unpublished.

\end{thebibliography}
\end{document}